\documentclass[prl,twocolumn,superscriptaddress,nofootinbib]{revtex4}
\usepackage{graphicx, graphics}
\usepackage{bm}
\usepackage{amssymb}
\usepackage{dcolumn}
\usepackage{color}
\usepackage{amsmath}
\usepackage{amstext}
\usepackage{ulem}
\usepackage{latexsym}
\usepackage{rotating}
\usepackage{tabularx}
\usepackage[usenames,dvipsnames]{xcolor}
\usepackage[colorlinks=true,citecolor=Cerulean,linkcolor=RubineRed,urlcolor=Cerulean]{hyperref}
\usepackage{tikz}

\begin{document}

\title{Quenched Dynamics of Artificial Spin Ice: Coarsening versus Kibble-Zurek}
 
\author{A. Lib\'{a}l}
\affiliation{Department of Mathematics and Computer Science, Babe{\c s}-Bolyai University, Cluj, Romania 400084}
\affiliation{Theoretical Division, Los Alamos National Laboratory, Los Alamos, NM 87545, USA}
\affiliation{Center for Nonlinear Studies, Los Alamos National Laboratory, Los Alamos, NM 87545, USA}

\author{A. del Campo}
\affiliation{Donostia International Physics Center,  E-20018 San Sebasti\'an, Spain}
\affiliation{IKERBASQUE, Basque Foundation for Science, E-48013 Bilbao, Spain}
\affiliation{Department of Physics, University of Massachusetts, Boston, MA 02125, USA}
\affiliation{Theoretical Division, Los Alamos National Laboratory, Los Alamos, NM 87545, USA}

\author{C. Nisoli}
\affiliation{Theoretical Division, Los Alamos National Laboratory, Los Alamos, NM 87545, USA}
\affiliation{Center for Nonlinear Studies, Los Alamos National Laboratory, Los Alamos, NM 87545, USA}

\author{C. Reichhardt}
\affiliation{Theoretical Division, Los Alamos National Laboratory, Los Alamos, NM 87545, USA}
\affiliation{Center for Nonlinear Studies, Los Alamos National Laboratory, Los Alamos, NM 87545, USA}

\author{C. J. O. Reichhardt}
\affiliation{Theoretical Division, Los Alamos National Laboratory, Los Alamos, NM 87545, USA}
\affiliation{Center for Nonlinear Studies, Los Alamos National Laboratory, Los Alamos, NM 87545, USA}

\date{\today}

\begin{abstract}
Artificial spin ices are ideal frustrated model systems in which to explore or design emergent phenomena with unprecedented characterization of the constituent degrees of freedom. In square spin ice, violations of the ice rule are topological excitations essential to the kinetics of the system, providing an ideal testbed for studying the dynamics of  such defects under varied quench rates. In this work we describe the first test of the Kibble-Zurek mechanism and critical coarsening in colloidal square and colloidal hexagonal ice under quenches from a  weakly interacting liquid state into a strongly interacting regime.  As expected, for infinitely slow quenches, the system is defect free, while for increasing quench rate, an increasing number of defects remain in the sample.  For square ice, we find regimes in which the defect population decreases as a power law with decreasing quench rate.  A detailed scaling analysis shows that for a wide range of parameters, including quench rates that are accessible by experiments, the behavior is described by critical coarsening rather than by the Kibble-Zurek mechanism, since the defect-defect interactions are long ranged.  For quenches closer to the critical point, however, there can be a competition between the two mechanisms.
\end{abstract}

\maketitle

\section{Introduction}
The term artificial spin ice (ASI)
describes a variety of
systems
that can be modeled by frustrated, interacting
effective
binary
degrees of freedom
which obey the ice rule.
The ASI size scales are
much larger than those of
real spin ice systems, allowing the individual spin degrees of
freedom to be imaged directly \cite{Wang06,Nisoli13,Gilbert16}.
ASI can be realized using arrays
of nanomagnets \cite{Wang06,Nisoli13,Gilbert16,Qi08,Ladak10,Mengotti11,Zhang13,Morgan11,Wang16},
colloids in ordered trap arrays \cite{Libal06,OrtizAmbriz16,Loehr16,Lee18,Libal18},
and vortices in nanostructured superconductors \cite{Libal09,Latimer13,Trastoy14,Xue17,Wang18}.  Of the
wide variety of different ASI geometries,
the first and most studied are
square~\cite{Wang06,Nisoli13,Gilbert16,Morgan11,Moller06} and hexagonal ices~\cite{Nisoli13,Qi08,Ladak10,Mengotti11,Zhang13,OrtizAmbriz16,Xue17,Libal18a}. While both obey the ice rule  in their low energy states,  the square geometry
produces an antiferromagnetic ground state, while the ice-manifold of hexagonal ice can remain disordered.

\begin{figure}
\includegraphics[width=3.3in]{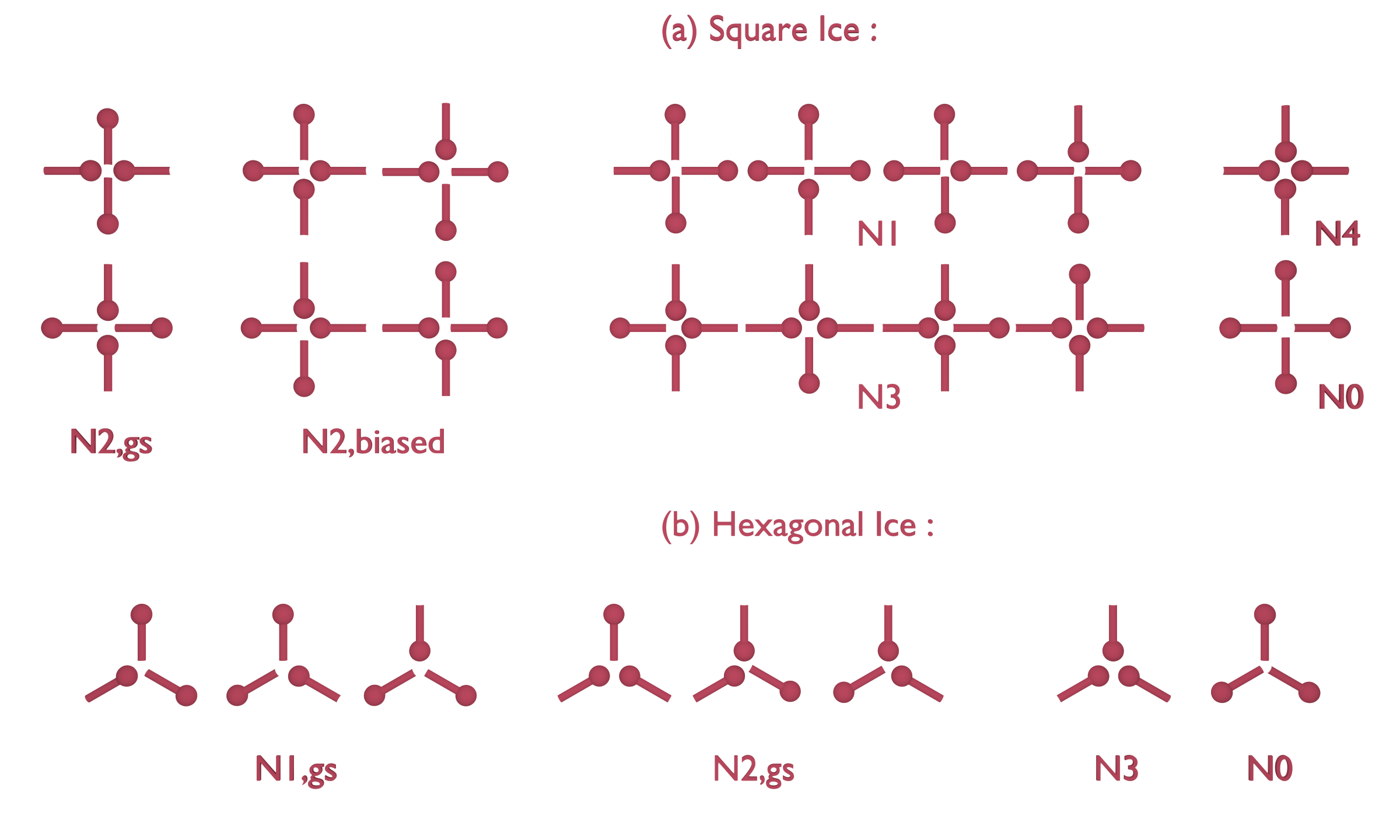}
\caption{(a) Vertex types for square ice. (b) Vertex types for hexagonal ice.  Dots indicate
the location of the particle with respect to the vertex.}
  \label{fig:table}
  \end{figure}

A particularly appealing 
feature of ASI systems
is that they contain well defined defects
that take the form of non-ice rule obeying vertices.
The system
can be characterized by
its different vertex types,
which can be labeled according to
the number
of spins pointing toward a vertex.
In the square ice, the vertices are named
$N_{n}$ where $n$ is
the number of spins pointing toward the vertex.
Here, $N_{0}$ and
$N_{4}$ are called double monopoles,
$N_{1}$ and $N_{3}$ are monopoles,
and the $N_{2, {\rm biased}}$ and $N_{2, {\rm gs}}$
are ice rule obeying vertices, where the latter is 
the ground state vertex configuration \cite{Nisoli13}. 
In Figure~\ref{fig:table}(a) we highlight the different vertex types for the  
square ice,  while Figure~\ref{fig:table}(b) shows the same for the honeycomb ice. 

At high temperatures or when the interactions between
neighboring effective spins are weak,
the two-dimensional square ice 
forms a liquid state with finite
non-ice rule obeying vertex populations.
As the temperature decreases or the interaction strength
increases,
there is a phase transition to  
a long range ordered state in which
only $N_{2, {\rm gs}}$ vertices are present \cite{Nisoli13}. 
In the square ice system there is a underlying second-order phase transition from the
ordered state to the disordered state \cite{Sendetsky2019} while for the
honeycomb ice this is 
only a crossover \cite{Nisoli13}. 

Previous work on ASI has generally focused on equilibrium states; however,
ASI are ideal systems in which to address current issues in 
nonequilibrium statistical mechanics.
For example,
the density of topological defects   
in a system after it is quenched at different rates
through a second order phase transition 
has implications for defect formation in the
early universe \cite{Kibble76,Zurek85},
vortex formation at normal to superconducting transitions \cite{Monaco09},
liquid crystal systems \cite{Bowick94},
Bose-Einstein condensates \cite{Weiler08,Lamporesi13,Navon15}, ion crystals \cite{Ulm13,Pyka13}, 
and manganites \cite{Griffin12}.

One scenario describing the behavior of the defects
for varied quenched rates is the
Kibble-Zurek (KZ) mechanism  
\cite{Kibble76,Zurek85,delCampo14},
in which the defect density $\rho_d$ increases with higher quench rate
according to a universal power law,
$\rho_{d} \propto \tau_{Q}^{-\beta}$,
where $\tau_{Q}$ is the
 the time duration of the quench or inverse quench rate.
Thus for large $\tau_{Q}$ or
slow quench rates, $\rho_{d}$ is expected to be small.
In the KZ mechanism, 
$\beta$
is related to the critical exponents
associated with the underlying equilibrium second order phase transition
through which the
system is quenched.
The KZ mechanism
relies on the adiabatic-impulse approximation according to which defects are produced when the system falls out of equilibrium (the freeze-out time scale).
In addition, it assumes that
the
density of defects arises exclusively from the nonadiabatic
crossing of the critical point, in the absence of any
dynamics in the ordered phase that may alter the defect population.
Other scenarios for
how the defect density could behave include a coarsening process
produced by the
motion and annihilation of defects
on the ordered side of the phase transition due to strong
defect-defect interactions \cite{Biroli10}. 
An ASI system is ideal for testing these different scenarios 
since excitations such as monopoles are very well defined and the universality
class of the phase transition
in many types of ASI, including the square ice, is known.
In addition,
since the square ice exhibits a phase transition but
the honeycomb ice
does not, the two types of ice should have
very different behaviors during a quench. 

Here we consider simulations and scaling analysis of a magnetically interacting
colloidal artificial spin ice.
The advantage of colloidal ice is 
that the strength of the colloid-colloid interactions 
can be tuned experimentally
as a function of time,
bringing the system from a non-interacting regime 
to a strongly interacting regime as a function of magnetic field and giving access to
a range of different quench rates.
We start the system in the weakly interacting disordered regime,
increase the magnetic field through the phase transition,
and measure the population
of the different vertex types as well as the
spatial configurations of the defects.
We consider both
square ice, where there is a
second order phase transition to
an ordered
ground state,
and hexagonal ice, where there is
only a crossover to a
disordered ice rule obeying state.
Our simulation faithfully mimics the experimental set up
as described in Refs.~\cite{OrtizAmbriz16,Loehr16,Libal17,Libal18}.
An advantage to studying a particle based model is that
the time-dependent
dynamics during the quench can be
directly accessed using
molecular dynamics techniques,
avoiding the issues that
arise
in using Monte Carlo (MC) methods to examine 
KZ scaling.
Different MC methods produce different 
results \cite{Liu2014},
while the MD approach faithfully represents the dynamics that
actually occur.

\begin{figure}
  \includegraphics[width=3.3in]{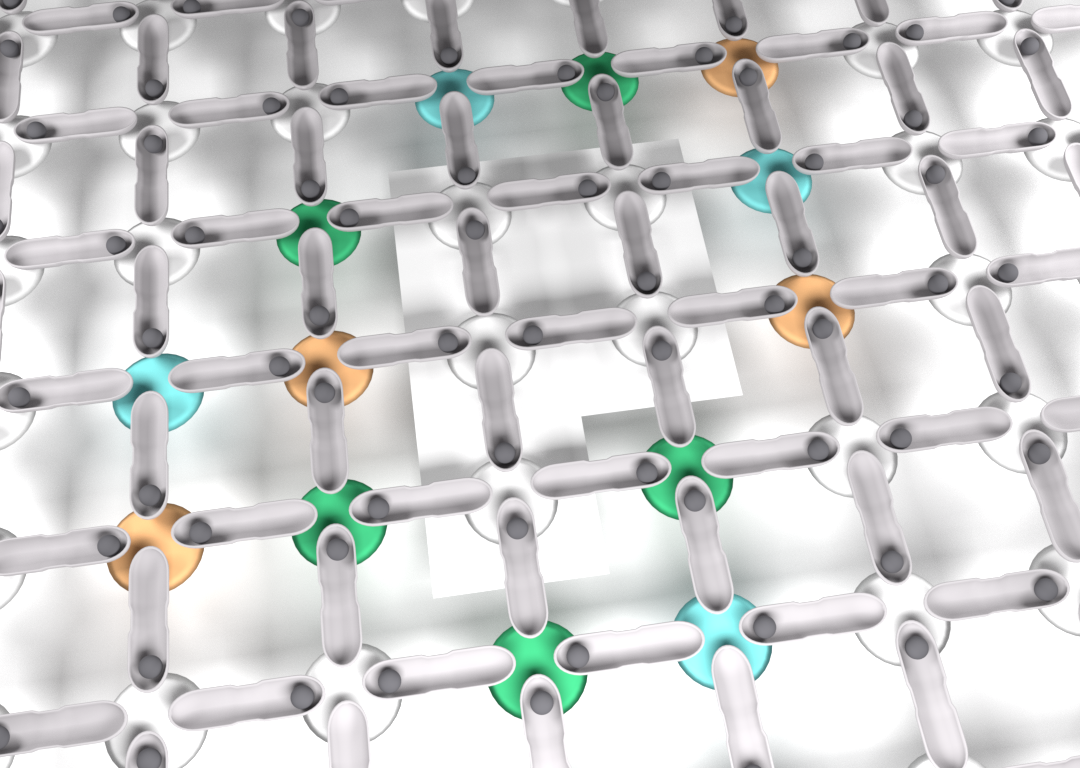}
\caption{Schematic of the square ice system.  Lozenges are the double well
traps which are each occupied by a single colloid, shown as a gray sphere,
that preferentially sits at one of the two ends of the trap.
In experimental realizations, the colloidal particles
are paramagnetic and repel each other with a strength that
can be controlled using an applied magnetic field.
The circles underneath each vertex are colored according to the vertex
type.
$N_{2, \rm{gs}}$ (the ground state): white.
$N_{2, \rm{biased}}$: green.
$N_1$: light blue.
$N_3$: orange.
Here there are no highly unfavorable
$N_0$ or $N_4$ vertices.
At the center of the image is a ground state cluster of vertices surrounded by
a grain boundary which separates it from vertices in a ground state with the
opposite orientation.}
\label{fig:1}
\end{figure}

\begin{figure}
  \includegraphics[width=3.3in]{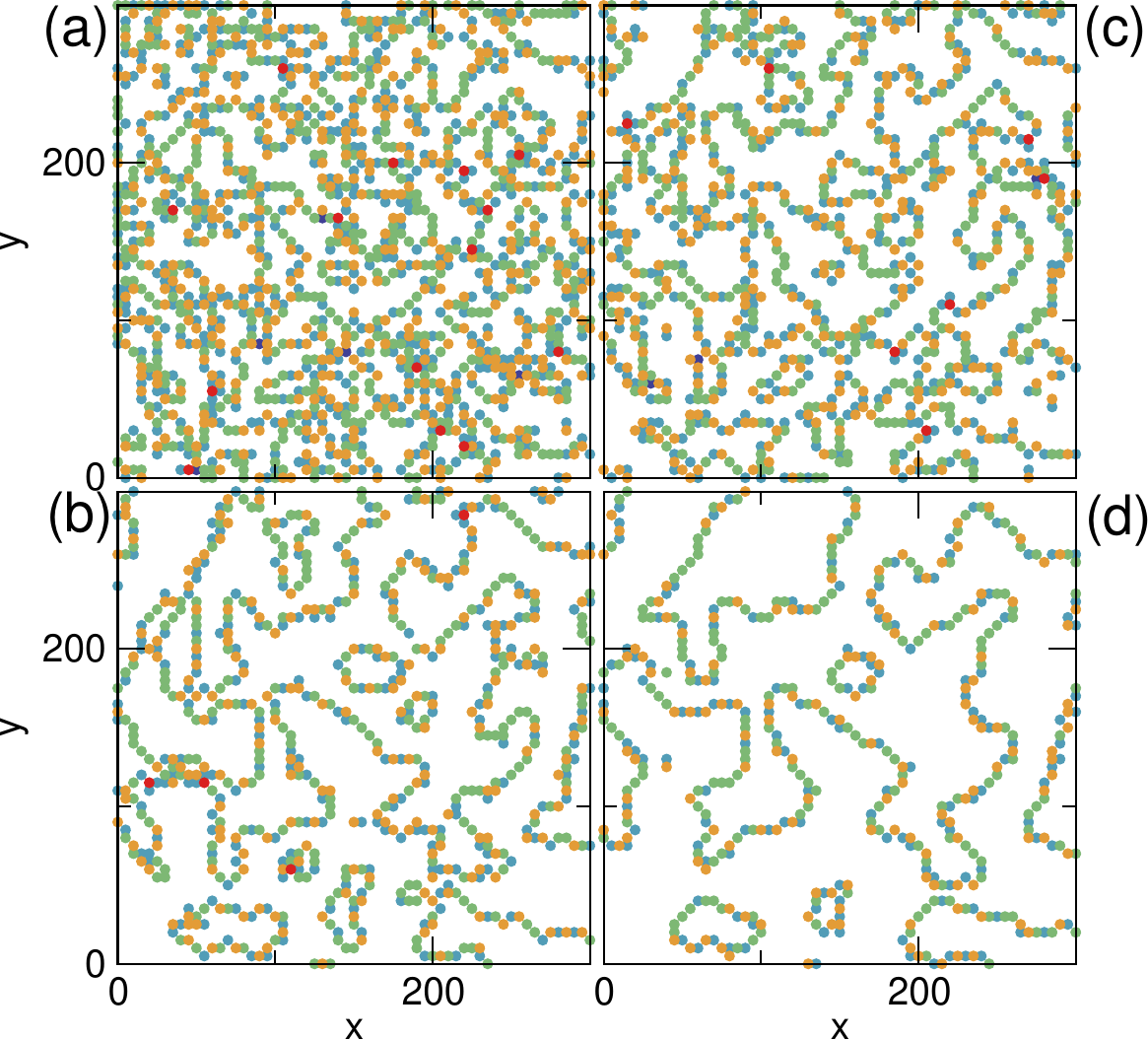}
\caption{Coarsening in the square ice system. Snapshots of a smaller part of the system for $\tau_Q = 80$ s total time.  a) $B = 16$ mT.
  b)  $B = 20$ mT.
  c) $B = 24$ mT.
  d) $B = 30$ mT.
  Dark blue and red dots: $N_0$ and $N_4$ vertices
  (double monopoles).
  Blue and orange dots:
  $N_1$ and $N_3$ vertices
  (monopoles).
  Green dots:
  $N_{2, {\rm biased}}$ vertices.
  The white areas contain $N_{2, {\rm gs}}$ ice ground state vertices.
}
\label{fig:2}
\end{figure}

\section{Results}

In Figure \ref{fig:1} we show a schematic of the
magnetically interacting colloids 
in a square ASI of 
double well traps.  Each elongated trap holds a single colloid which can sit on
either end of the trap,
determining the direction of the effective spin.   
The colloid-colloid interaction force is given by
$F_{\mathrm{pp}}(r)=A_{\mathrm{c}}/r^4$ with  $A_{\mathrm{c}}=3\times 10^6\chi_m^2 V^2 2B^2/(2\pi\mu)$ 
for particles a distance $r$ apart, where $B$ is the magnetic field. 
For our parameters, the critical
magnetic field at which the equilibrium system 
orders into a defect free ground state is $B_c = 9$ mT. 
We start the system at $B = 0.0$ and increase
the field to $B=40$ mT
at different sweep rates.
Figure \ref{fig:2} shows the vertex populations with 
the same color scheme from Figure \ref{fig:1}
in a simulation with a
quench time duration of $\tau_{Q} = 80$ s
at several values of $B$.
The defects form closed loop
grain boundaries
similar to those observed in
square ice systems with
varied amounts of quenched disorder \cite{Morgan11,Libal09,Budrikis12,Drisko17}. 
For faster quench rates or smaller $\tau_{Q}$, the number of
non-ground state vertices increases
and the grain boundaries are smaller.

\begin{figure}
  \includegraphics[width=3.3in]{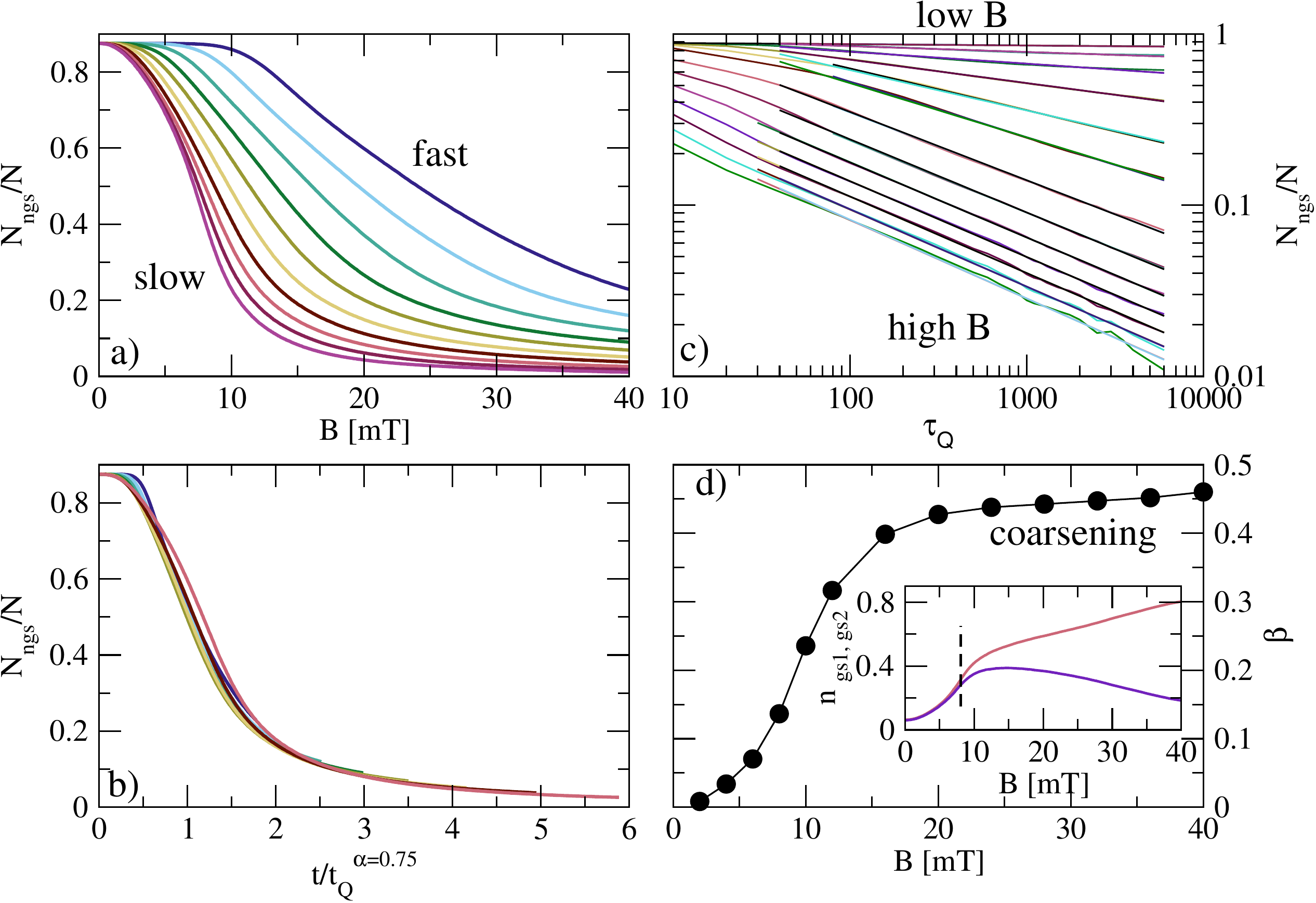}
\caption{Transition from the ordered to the disordered state as a function of quenching speed.
  a) Fraction of non-ground state vertex types
  $N_{\rm ngs}/N$ in the system vs magnetic field values $B$.
  From dark blue to dark red, the curves represent total run times of $\tau_Q=10$,
  20, 40, 80, 150, 300, 600, 1200, 2500, and $6000$ s.
  b) Rescaling of
  $N_{\rm ngs}/N$
  vs
  $t/\tau_Q^{\alpha}$, the time
divided by the total quench time raised to the power $\alpha=0.75$.
c) Scaling of
$N_{\rm ngs}/N$
as a function of 
quenching time
$\tau_Q$
for different magnetic field values.
From dark blue to dark red, the curves represent constant magnetic fields of $B=4$ mT (dark blue), $B=6$ mT (light blue) $B=8$ mT (light green), and subsequently, $B=10$,
12, 16, 20, 24, 28, 32, 36, and 40 mT.
d) Power law exponents $\beta$ obtained from
the data in panel c) vs the magnetic field value $B$.
Inset: The fraction of the larger ground state cluster $n_{\rm gs1}=N_{\rm gs1}/N$ (upper pink
line) and the fraction of the smaller ground state cluster $n_{\rm gs2}=N_{\rm gs2}/N$ (lower
purple line) as a function of $B$, showing a bifurcation at the critical field (dashed line),
corresponding to the spontaneous symmetry breaking.
}
\label{fig:3}
\end{figure}

In Fig.~\ref{fig:3}(a) we plot the fraction of 
non-ground state vertices $N_{\rm ngs}/N$ versus $B$ at different sweep rates.
The fastest transition with $t=10$ s 
is denoted by the rightmost blue line, and the quench rate decreases for curves
that are further to the left.
The systems are initialized
in a completely random configuration at $B=0.0$ with
$N_{\rm ngs}/N=7/8$.
As the quench rate decreases,
the value of $N_{\rm ngs}/N$ decreases. 
In Figure~\ref{fig:3}(b) we show that the
$N_{\rm ngs}/N$ curves from Figure~\ref{fig:3}(a) can be collapsed
by dividing the time by $\tau_Q^\alpha$, where $\alpha=3/4$. 

In Figure~\ref{fig:3}(c)
we plot $N_{\rm ngs}/N$ versus the quench time $\tau_Q$ at
different fixed values of the magnetic field
from $B=4$ mT (top) to $B=40$ mT (bottom).
The runs were performed over the experimentally accessible range of
$t=10$ s to $t=6000$ s.
We fit each curve to 
a power law with
$N_{\rm ngs}/N \propto \tau_{Q}^{-\beta}$,
where $N_{\rm ngs}/N=\rho_d$, and we
plot the resulting exponents $\beta$ versus $B$ in
Figure~\ref{fig:3}(d).
For $B < 9$ mT,
the system does not order at all, 
while for $B >12$ mT, the exponent saturates at
$\beta= 0.45$.
This indicates that we have
two different regimes of behavior.
For smaller magnetic fields between the values 
of $B=10$ mT and $B=12$ mT,
we find a slower decay rate with an exponent between
$\beta=0.2$ and $\beta = 0.3$.

\subsection{Kibble-Zurek Mechanism}
Now that we have established that our system has
both a critical point and power law scaling
of the defect density for different quench rates,
we can test whether our results
are consistent with the KZ mechanism \cite{Kibble76,Zurek85,delCampo14}.
In particular, the lag time between the nonequilibrium and
equilibrium value scale is
expected to be set by the so-called freeze-out time
$\hat{t} \sim (\tau_0 \tau_Q^{z\nu})^{\frac{1}{1+z\nu}} \sim \tau_Q^{\frac{z\nu}{1+z\nu}}$.

To investigate whether the transition obeys the KZ mechanism,
we collapse the runs with different quench times together by rescaling the time axis. 
In Figure \ref{fig:3}(b) we show
the evolution of $N_{\rm ngs}/N$ versus time where the time has been
divided
by a power of the quench time $\tau_Q^\alpha$  .
The collapse is achieved with $\alpha=3/4$.
The KZ mechanism prediction then implies 
that ${\frac{z\nu}{1+z\nu}}=\alpha=3/4$; however, the square ice
falls into the Ising universality class with $\nu = 1$ and $z = 2$ \cite{Wu69}, which
gives ${\frac{z\nu}{1+z\nu}}=2/3$.    

Another prediction of the KZ mechanism is that the total 
number of defects scales as
$\rho_{d}\sim\tau_Q^{-\frac{D\nu}{1+z\nu}}$, where 
$D$ is the dimension of the system.  In our case, $D=2$
and $\rho_d=N_{\rm ngs}/N$.
The 2D Ising model  
gives a prediction of  $\frac{D\nu}{1+z\nu}=2/3$,
but in Fig.~\ref{fig:3}(c) we find 
$\rho_{d}\sim \tau_{Q}^{-1/2}$
or $\frac{D\nu}{1+z\nu}=1/2$, indicating that
the scaling of the defects that we obtain
is not a result of the KZ mechanism.
For quenches out to higher values of $B$, the defects
such as $+1$ and $-1$ monopoles are strongly 
interacting and 
undergo a nonnegligible amount of dynamical motion
via their
effective Coulomb interactions,
as has been observed in colloidal experiments \cite{Loehr16} 
and simulations \cite{Libal17}. 
The presence of defect dynamics
during the part of the quench in the ordered state
violates one of the assumptions for the KZ scenario.  
We note that for coarsening dynamics
near a critical point,
the ordered regions of radius $R$ grows as
$R(t) \propto t^{1/z}$ \cite{Hohenberg77}, which for the Ising model gives
$R(t) \propto t^{1/2}$,
where $t$ is time.
If the size of the ordered regions
grows, the number of defects 
could be proportional to $1/R(t)$, in agreement with our observations.
In other types of ASI, such as nanomagnetic systems, it is possible
that the KZ regime could be
accessed more easily since
the motion of the defects is slower.
Alternatively, there could be a regime of KZ behavior at much faster
quench rates than those we considered, for which
the defects simply do not have time to move.

We note that although,
up to a nearest neighbor approximation, a {\it magnetic} square ice can be mapped into a $J_1, J_2$ Ising system~\cite{wu1969critical}, the colloidal square ice differs greatly from
the magnetic square ice, both in energetics and in the nature of its frustration~\cite{nisoli2014dumping,Libal18}. The colloidal square ice can only be mapped
exactly into a magnetic square ice at equilibrium~\cite{Nisoli18,Levis13,Sendetsky2019}.
This is because
the colloidal ice contains many more states, corresponding to colloids in between preferential positions, which might make its out-of-equilibrium kinetics much different from
those of its magnetic analogue. An example of the difference
between these two ice systems appears in Ref.~\cite{Libal17}.

\subsection{Hexagonal system}

\begin{figure}
  \includegraphics[width=3.3in]{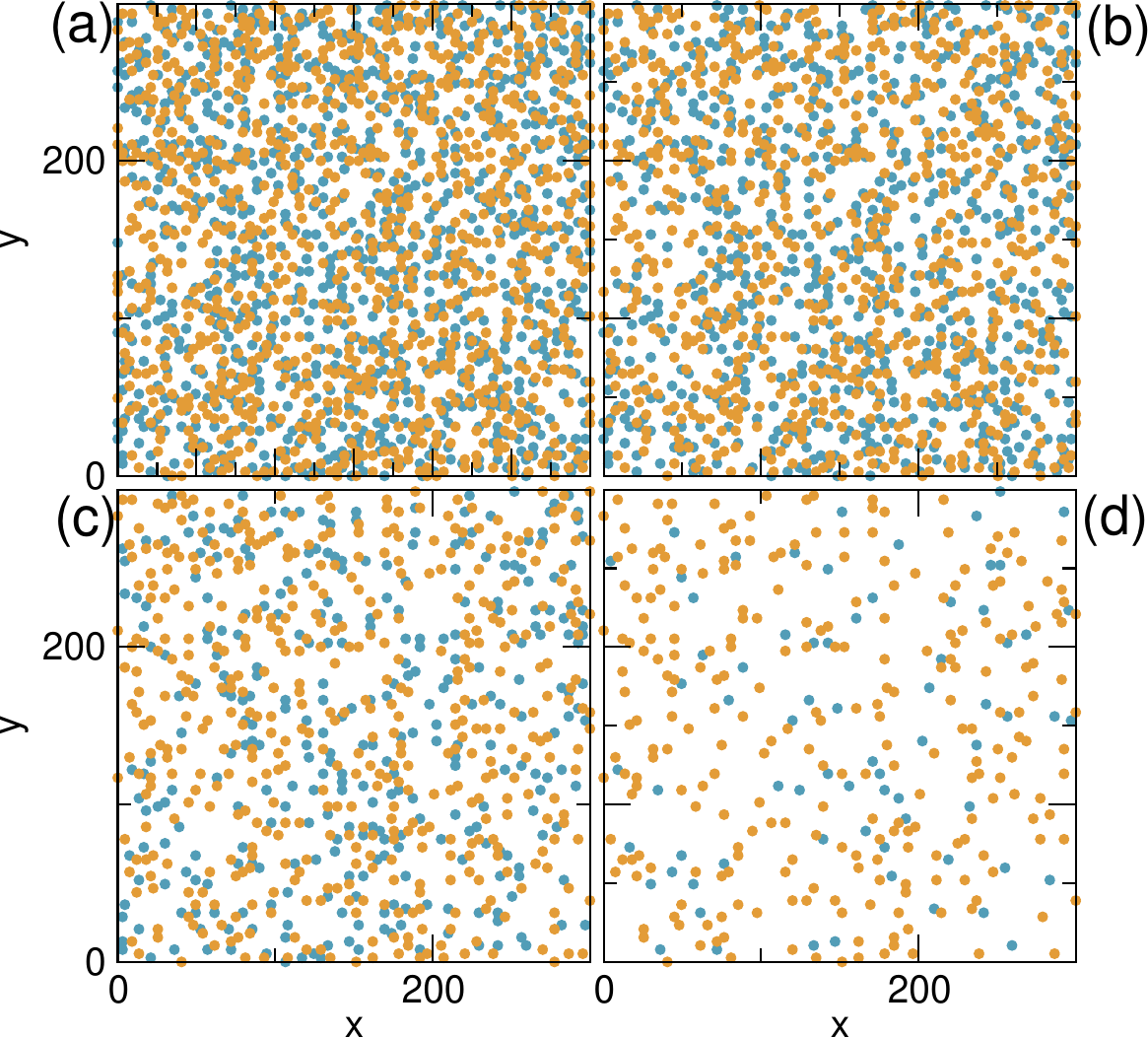}
\caption{ Transition in the hexagonal ASI.
  Snapshots of a small part of the system for $\tau_Q = 80$ s
  total time.  a) $B = 8$ mT.  b)  $B = 10$ mT. c) $B = 12$ mT.
  d) $B = 14$ mT.
  Blue dots: three-out vertices; orange dots: three-in vertices.
  Ground state vertices in the white regions are not plotted.}
\label{fig:6}
\end{figure}

In the hexagonal ASI,
each vertex is surrounded by 
only three elongated pinning sites.
Unlike the square ice,
the hexagonal ASI has
no phase transition from a disordered to an ordered 
phase, so we would not expect the KZ scenario to apply.
We conduct the same type of simulation from 
a zero field state to a maximum
field of $B=40$ mT, where the equilibrium configurations
at higher $B$ do not contain any monopoles.
In Figure  \ref{fig:6} we show snapshots of the
transition in the hexagonal ASI as a 
function of increasing interaction strength
$B=8$, 10, 12, and 14 mT for a quench time of $\tau_Q=80$ s. 
In this case, non-favorable vertex types disappear
during the crossover to the disordered ice-rule obeying state
without forming any spatially correlated structures or grain boundaries
of the type observed in the square ice system.
Therefore, the defect dynamics and coarsening should be different between
the two systems.
In the initial random configuration,
the ground state vertices
in the hexagonal ice already occupy $N_{\rm gs}/N=3/4$ of the system,
in contrast to the square system where $N_{\rm gs}/N=1/8$ at initialization.
As a result, the hexagonal ice does not need to nucleate and grow clusters of
ice rule obeying vertices.

\begin{figure}
  \includegraphics[width=3.3in]{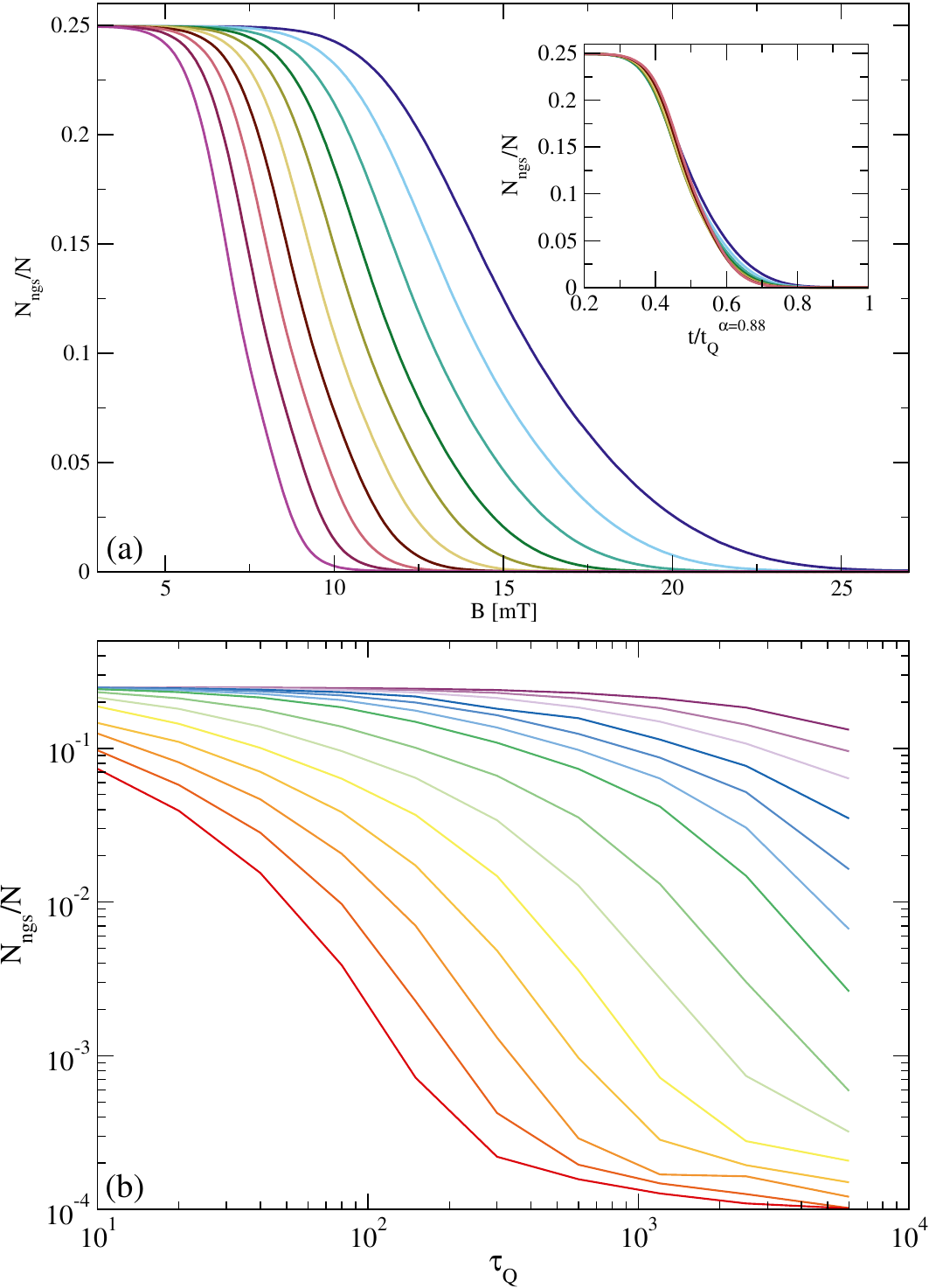}
\caption{(a) Measure of the transition in the hexagonal ASI system.
  Non-ground state vertex fraction $N_{\rm ngs}/N$ as a function
  of applied field for $\tau_Q=10$, 20, 40, 80, 150, 300, 600, 1200, 2500,
  and 6000 s, from blue (right) to red (left). Inset: rescaling of
  $N_{\rm ngs}/N$ as a function of time divided by $\tau_Q^\alpha$, where $\alpha=0.88$.
  (b) A log-log plot of $N_{\rm ngs}/N$ vs $\tau_{Q}$ for the system in
  panel (a) at $B = 7.0$, 7.5, 8,  8.5, 9, 9.5, 10, 11, 12, 13, 14,
  15, 16, and $17$ mT, from purple (top) to red (bottom).
  showing the lack of a power law decay of defects, in contrast to the square ice case. }
\label{fig:7}
\end{figure}

In Figure \ref{fig:7} we plot the number of non-ground state vertices as a function of 
the applied field for different $\tau_Q $ values
ranging from $\tau_Q=10$ s to $\tau_Q=6000$ s
for the hexagonal ice. 
The transition happens
over a narrower range of $B$ than in the square ice
since there are no kinetic barriers to overcome in the process of eliminating the 
non-ground state defects.
This
also gives a higher exponent of
$\alpha=0.88$
in the rescaling of $N_{\rm ngs}/N$ versus
$t/\tau_{Q}^{\alpha}$
shown in the inset of Fig.~\ref{fig:7},
indicating that the defect annihilation mechanism differs from what
is found in the square ice system.
In the square ice,
the monopoles are located along
grain boundaries
and annihilate as the grain boundaries shrink.
In contrast, the monopoles in the hexagonal ice
are 
not on grain boundaries and
can move toward each other
along straight paths, as found in experiments \cite{Mengotti11}.
  More relevantly, monopoles in hexagonal ice are not topologically protected. While charges $\pm 2$ of square ice cannot be reabsorbed but can only be annihilated and created in pairs, $\pm 3$ charged violations of the ice rule in hexagonal ice can appear and disappear individually. That is because even ice rule vertices are charged ($\pm1$) and thus each monopole in hexagonal ice can transfer charge to the surrounding plasma.

In Fig.~\ref{fig:7}(b), the
plot of defect density $N_{\rm ngs}/N$ versus the quench duration $\tau_Q$
for the hexagonal ice at varied $B$
shows that there is no power law behavior
in the density of defects, which is consistent with the lack of a phase transition in  the system.
As a result,
the KZ mechanism scenario
does not apply,
and critical coarsening cannot occur due to the lack of a critical
point.   

\subsection{ Discussion}

Our results can be compared directly to current
experimental colloidal ASI systems.
The experimental samples
are smaller than what is considered in our simulations;
however, the experiments could be
repeated many times to
improve the statistics.
In magnetic ASI, there is now a system
in which thermal
transitions from a liquid to an ordered state can be realized
\cite{Nisoli13,Kapaklis14,Anghinolfi15}.
In such samples, 
quenches can be performed
by varying the rate
at which the temperature is
swept across the transition from the liquid to ordered state.
In superconducting systems, where
artificial ices can be realized using magnetic flux lines,
a similar temperature control could be used at finite fields in 
passing from a normal to a superconducting state as a function of temperature.
It is possible
that the coarsening dynamics in the magnetic AFI
could differ from that found in the particle-based AFI
since the particle based system
minimizes the global energy rather than the vertex energy, making the resulting
ice state
more fragile \cite{Libal18,Nisoli18}.
The kinetics of annihilation and spin flipping are
also likely to depend on the microscopic details of the particular
ASI realization.
Many ASI systems
have Coulomb interactions between the monopoles, and in these it is
possible that the KZ mechanism
always
competes with coarsening.
One possible experiment to test this 
would be to create magnetic nanoislands
that are sufficiently far apart
to reduce the strength of the defect-defect interactions and
minimize the coarsening.
Other future directions are
to consider alternative ASI geometries
\cite{Nisoli13,Wang16,Libal18,Farhan17,Lao18,Sklenar19}, including geometries
in which the monopoles are not as strongly bound \cite{Perrin16,Farhan19}  
It would also be interesting to study the effect
of disorder to see if the exponents change or whether
 glassy dynamics arise such as a crossover to a logarithmic
 rather than a power law decay.
 It may also be possible that a small amount of disorder could slow down the
dynamics of the defects enough that the KZ mechanism regime could be accessed.

\subsection{Conclusion}  
In conclusion, we have examined the defect density populations for varied quench
rates from a disordered to an ordered state in
square and
hexagonal magnetically interacting 
colloidal spin ice systems.
In the square ice,
we find that when the quench
into the ordered state is sufficiently deep,
there is a power law decay of the defect density
with
$\rho_{d} \propto \tau_Q^{-1/2}$. 
  Based on scaling arguments 
for the university class
of the square ice, we find that the
behavior of the quenched square ice is governed by coarsening rather than the Kibble-Zurek mechanism.
The lack of KZ behavior is
likely due to the strong Coulomb
interactions between the monopoles.
This causes a considerable amount of defect dynamics to occur
during the quench,
while the KZ mechanism assumes that no dynamics occurs in the ordered phase.
In the case of the hexagonal ice, which has no second order phase transition to
an ordered state,
we find a
very different type of defect configurations
as well as a lack of power law scaling of the defect density with 
varied quench rates.  

Our results could be compared with quenches of
different types of AFI in magnetic, colloidal, and superconducting
systems.
Each of these systems could exhibit
different interactions between the defects or
different
kinetics, and it is possible that one or more of the systems could have
a regime in which the KZ mechanism is observable.

\section{Methods}

We simulate a system of colloidal superparamagnetic particles with a radius of $r=1\mu$m. The particles are placed in a square $100 \times 100$ lattice containing
$20,000$ particles and $10,000$ pinning sites
or in a hexagonal $38 \times 66$ lattice containing
$15,048$ particles and $10,032$ pinning sites.
Each pinning site is a double-well trap and the
lattice constant is $a_x = a_y = 5.0 \mu$m in the square
ice lattice and  $a_x= 3 \mu$m and $a_y = 3\sqrt{3}/2\mu$m in the hexagonal
ice lattice,
giving a total system size of $500 \times 500 \mu$m for the square ice
and $342 \times 342.95 \mu$m for the hexagonal ice.
We use periodic boundary  conditions in both the $x$ and $y$ directions.

The elongated gravitational double-well traps are modeled as two spherical quarters connected by an elongated half-cylindrical trough of length $2\mu$m in the square ice
and $1.4\mu$m in the hexagonal ice that has a repulsive bump in the middle.
Each minimum of the double well is located at the end of the elongated
trough, coinciding with the minimum in the spherical quarter.
When the particle is in either of the spherical ends, an unbreakable harmonic spring force tethers the particle to the minimum with a spring constant of $k = 0.222$pN/$\mu$m for the square ice and $k= 2.22$ pN/$\mu$m for the hexagonal ice.
When the particle is in the elongated part of the pin,
the same unbreakable harmonic spring force acts on it in the direction perpendicular to the elongated trough,
and an additional force is exerted by the bump in the middle of the
trough which has a maximum value of
$F_{sm} =0.011$ pN for the square ice and $F_{sm} =0.211$ pN for the hexagonal ice.
This force decays to zero linearly in each half of the elongated trough as the
intersection with the spherical quarters is approached.
These forces together compose the substrate force denoted as $F_s^i$.

We use a smaller lattice constant for the less densely packed hexagonal ice because
stronger inter-particle interactions are required to induce the spin ice ordering
compared to the square ice system.
We also increase the pinning strength significantly for the hexagonal ice
to prevent the particles from
ordering into a triangular lattice with each particle sitting at the center of the
elongated trough,
which destroys the spin ice nature of the particle based system.
With the chosen values, which are within the experimentally realizable
regime, the spin ice manifold is preserved.

Magnetization of the particles in the $z$ direction by an external magnetic field produces a repulsive
particle-particle interaction force $F_{\mathrm{pp}}(r)=A_{\mathrm{c}}/r^4$ with  $A_{\mathrm{c}}=3\times 10^6\chi_m^2 V^2 2B^2/(2\pi\mu)$ for particles a distance $r$ apart. Here $\chi_m$ is the magnetic susceptibility, $V$ is the particle volume,
$B$ is the magnetic field in mT, and all distances are measured in $\mu$m.
At $B=40$ mT, the maximum field we consider, this gives $F_{\mathrm{pp}}=0.49$pN for $r=3\mu$m, which is a typical distance for the square ice,
and  $F_{\mathrm{pp}}=6.05$pN for $r=1.6\mu$m, which is a typical distance for the
hexagonal ice.
The dynamics of colloid $i$ are obtained using the following
discretized overdamped equation of motion:

\begin{equation}
 \frac{1}{\mu}\frac{\Delta {\bf r}_i}{\Delta t} = \sqrt{\frac{2}{D\Delta t}}k_{\mathrm{B}} T N[0,1] + F_{\mathrm{pp}}^i + F_{\mathrm{s}}^i
\end{equation}

Here the  diffusion constant $D=36000$ nm$^2$/s, the mobility $\mu = 8.894 \mu$m/s/pN  and the simulation time step $\Delta t = 1$ms. The first term on the right is a  thermal force consisting of Langevin kicks of magnitude $F_T=0.954$ pN corresponding to a temperature of $t=20^\circ$C.
Here, $N[0,1]$ denotes a random number drawn from a normal (Gaussian) distribution with a mean of $0$ and a standard deviation of $1$.
Each trap is filled with a single particle which is randomly placed in one of the two
minima.
We increase $B$ linearly from $B = 0$ mT to $B = 40$ mT, following a procedure that is feasible to achieve experimentally. We average the results over $100$ simulations performed with different random seeds.

{\it Acknowledgements---}  
We gratefully acknowledge the support of the U.S. Department of
Energy through the LANL/LDRD program for this work.
This work was supported by the US Department of Energy through
the Los Alamos National Laboratory.  Los Alamos National Laboratory is
operated by Triad National Security, LLC, for the National Nuclear Security
Administration of the U. S. Department of Energy (Contract No. 892333218NCA000001).

\bibliography{mybib}

\end{document}